\documentclass{aa}  
\usepackage{graphicx}
\usepackage{txfonts}
\usepackage[table,xcdraw]{xcolor}

\begin{document} 
   \title{On the faintest solar coronal hard X-rays observed with FOXSI}
  \author{Juan Camilo Buitrago-Casas
          \inst{1,2},
          Lindsay Glesener
          \inst{3},
          Steven Christe
          \inst{4},
          S\"am Krucker
          \inst{1,5},
          Juliana Vievering
          \inst{3,6},
          P.S. Athiray
          \inst{3,7},
          Sophie Musset
          \inst{3,8},
          Lance Davis
          \inst{3},
          Sasha Courtade
          \inst{1},
          Gregory Dalton
          \inst{1},
          Paul Turin
          \inst{9},
          Zoe Turin
          \inst{10},
          Brian Ramsey
          \inst{7},
          Stephen Bongiorno
          \inst{7},
          Daniel Ryan
          \inst{4,5},
          Tadayuki Takahashi
          \inst{11,12},
          Kento Furukawa
          \inst{12},
          Shin Watanabe
          \inst{13,11},
          Noriyuki Narukage
          \inst{14},
          Shin-nosuke Ishikawa
          \inst{15},
          Ikuyuki Mitsuishi
          \inst{16},
          Kouichi Hagino
          \inst{17},
          Van Shourt
          \inst{1},
          Jessie Duncan,
          \inst{3},
          Yixian Zhang,
          \inst{3},
          \and
          Stuart D. Bale
          \inst{1,2}
          }
  \institute{Space Sciences Laboratory, University of California Berkeley,                Berkeley, CA, USA\\
              \email{milo@ssl.berkeley.edu}
         \and
              Physics Department, University of California, Berkeley, CA, USA
         \and
              University of Minnesota, 
              Physics \& Astronomy, Minneapolis, MN, USA
         \and
              NASA Goddard Space Flight Center, Greenbelt, MD, USA
         \and
              University of Applied Sciences and Arts Northwestern Switzerland, Windisch, Switzerland
         \and 
              Johns Hopkins University Applied Physics Laboratory, Laurel, MD, USA
         \and
              NASA Marshall Space Flight Center, Huntsville, AL, USA
         \and
              ESA, European Space Research and Technology Centre (ESTEC), The Netherlands
         \and
              Heliospace Corporation, Berkeley, CA, USA
         \and
              University of Colorado, Boulder, CO, USA
         \and
              Kavli Institute for the Physics and Mathematics of the Universe (WPI) , The University of Tokyo, Japan
         \and
              Department of Physics, The University of Tokyo, Japan
         \and
              Institute of Space and Astronautical Science, Kanagawa, Japan
         \and
              National Astronomical Observatory of Japan, Tokyo, Japan
         \and
              Rikkyo University, Graduate School of Artificial Intelligence and Science, Tokyo, Japan
         \and
              Nagoya University, Graduate School of Science, Aichi, Japan 
         \and
              Tokyo University of Science, Shinjuku City, Tokyo, Japan
             }


  \abstract
   {Solar nanoflares are small impulsive events releasing magnetic energy in the corona. If nanoflares follow the same physics as their larger counterparts, they should emit hard X-rays (HXRs) but with a rather faint intensity. A copious and continuous presence of nanoflares would result in a sustained HXR emission. These nanoflares could deliver enormous amounts of energy into the solar corona, possibly accounting for its high temperatures. To date, there has not been any direct observation of such persistent HXRs from the quiescent Sun. However, \citeauthor{hannah2010constraining} in \citeyear{hannah2010constraining} constrained the quiet Sun HXR emission using almost 12 days of quiescent solar-off-pointing observations by RHESSI. These observations set $2\sigma$ upper limits at $3.4\times 10^{-2}$ photons s$^{-1}$ cm$^{-2}$ keV$^{-1}$ and $9.5\times 10^{-4}$ photons s$^{-1}$ cm$^{-2}$ keV$^{-1}$ for the 3-6 keV and 6-12 keV energy ranges, respectively.}
   {Observing faint HXR emission is challenging because it demands high sensitivity and dynamic range instruments. The Focusing Optics X-ray Solar Imager (FOXSI) sounding rocket experiment excels in these two attributes when compared with RHESSI. FOXSI completed its second and third successful flights (FOXSI-2 and -3) on December 11, 2014, and September 7, 2018, respectively. This paper aims to constrain the quiet Sun emission in the 5-10 keV energy range using FOXSI-2 and -3 observations.}
   {To fully characterize the sensitivity of FOXSI, we assessed ghost ray backgrounds generated by sources outside of the field of view via a ray-tracing algorithm. We use a bayesian approach to provide upper thresholds of quiet Sun HXR emission and probability distributions for the expected flux when a quiet-Sun HXR source is assumed to exist.}
   {We found a FOXSI-2 upper limit of 4.5$\times 10^{-2}$ photons s$^{-1}$ cm$^{-2}$ keV$^{-1}$ with a $2\sigma$ confidence level in the 5-10 keV energy range. This limit is the first-ever quiet Sun upper threshold in HXR reported using $\sim$ 1-minute observations during a period of high solar activity. RHESSI was unable to measure the quiet Sun emission during active times due to its limited dynamic range. During FOXSI-3's flight, the Sun exhibited a fairly quiet configuration, displaying only one aged non-flaring active region. Using the entire $\sim$6.5 minutes of FOXSI-3 data, we report a $2\sigma$ upper limits of $\sim10^{-4}$ photons s$^{-1}$ cm$^{-2}$ keV$^{-1}$ for the 5-10 keV energy range.}
   {FOXSI-3's upper limits on quiet Sun emission are similar to that reported by  \cite{hannah2010constraining}, but FOXSI-3 achieved these results with only 5 minutes of observations or about 1/2600 less time than RHESSI. A possible future spacecraft using hard X-ray focusing optics like FOXSI's concept would allow enough observation time to constrain the current HXR quiet Sun limits further or perhaps even make direct detections. This is the first report of quiet Sun HXR limits from FOXSI and the first science paper using FOXSI-3 observations.}

               
   \authorrunning{J.C. Buitrago-Casas, et.al}
   \maketitle

\section{Introduction}

In solar and heliophysics, the coronal heating problem relates to the puzzle of identifying and understanding the mechanism(s) causing corona's temperatures to be multiple thousands times hotter than the solar surface \citep[e.g.,][]{klimchuk2006solving, klimchuk2015key}. Among the various plausible hypotheses proposed, the two strongest candidates are i) MHD wave dissipation and ii) copious low energy magnetic reconnections (or "nanoflares" as coined by \citeauthor{parker1988nanoflares} in \citeyear{parker1988nanoflares}) \citep[e.g.,][]{hudson1991solar,bogachev2020microflares}. \citet{klimchuk2006solving} pointed out that, when examined thoroughly, most plausible coronal heating explanations imply non-thermal heating that happens impulsively on individual flux tubes (strands). If such small, impulsive events follow the physics of larger flares, non-thermal electrons energized during the small, ubiquitous reconnections in the corona should be the base for heating the coronal plasma. The emission of hard X-rays (HXRs) is a direct consequence of these non-thermal electrons slowing down in the chromosphere. HXRs have been observed in non-flaring active regions, revealing the presence of hot plasma over 7 million Kelvin \citep[e.g., ][]{ishikawa2017detection}. Other authors have shown evidence of non-thermal particles in microflares (typical energies of $E\sim10^{27}$ erg) by directly analyzing their emission in HXRs \citep[e.g., ][]{christe2008rhessi,hannah2011microflares,glesener2020accelerated,duncan2021nustar}. For nanoflares, with energies of $E\sim10^{24}$ erg or less, HXRs are far fainter than those from larger flares and challenging to detect due to the limited sensitivity of current instruments. Other direct and indirect observations in different wavelengths can (so far) only be explained with the presence of non-thermal particles. Some instances are (a) UV IRIS spectral observations of short-lived brightenings at loop footpoints in non-flaring active regions \citep[e.g., ][]{testa2014evidence,testa2020iris}, (b) radio observations of non-thermal emission from the quiescent solar corona \citep[e.g., ][]{james2018energetics,mondal2020first}, and (c) spectral observations suggesting non-Maxwellian distributions \citep[e.g., ][]{dudik2017non,dudik2017nonequilibrium}. Nanoflare observations in HXRs can complement our current understanding of the non-thermal particle processes' role in heating the quiescent corona.


In recent years, the Focusing Optics X-ray Solar Imager (FOXSI) sounding rocket experiment has provided high sensitivity (and high dynamic range) solar X-ray observations in the band of $\sim$4-20 keV, with capabilities to perform imaging spectroscopy at 8.8 arcsec spatial and 0.5 keV energy resolutions \citep{krucker2014first,christe2016foxsi,musset2019ghost}. FOXSI observed areas in the solar disk free of active regions during its second and third flights. Analyzing the very few counts observed with FOXSI when pointing to the quiet Sun is currently the best way we have to evaluate the faintest sources of HXRs from the solar corona, which is the core of this work.


\section{The FOXSI sounding rocket}

\begin{figure}
\centering
\includegraphics[width=\hsize]{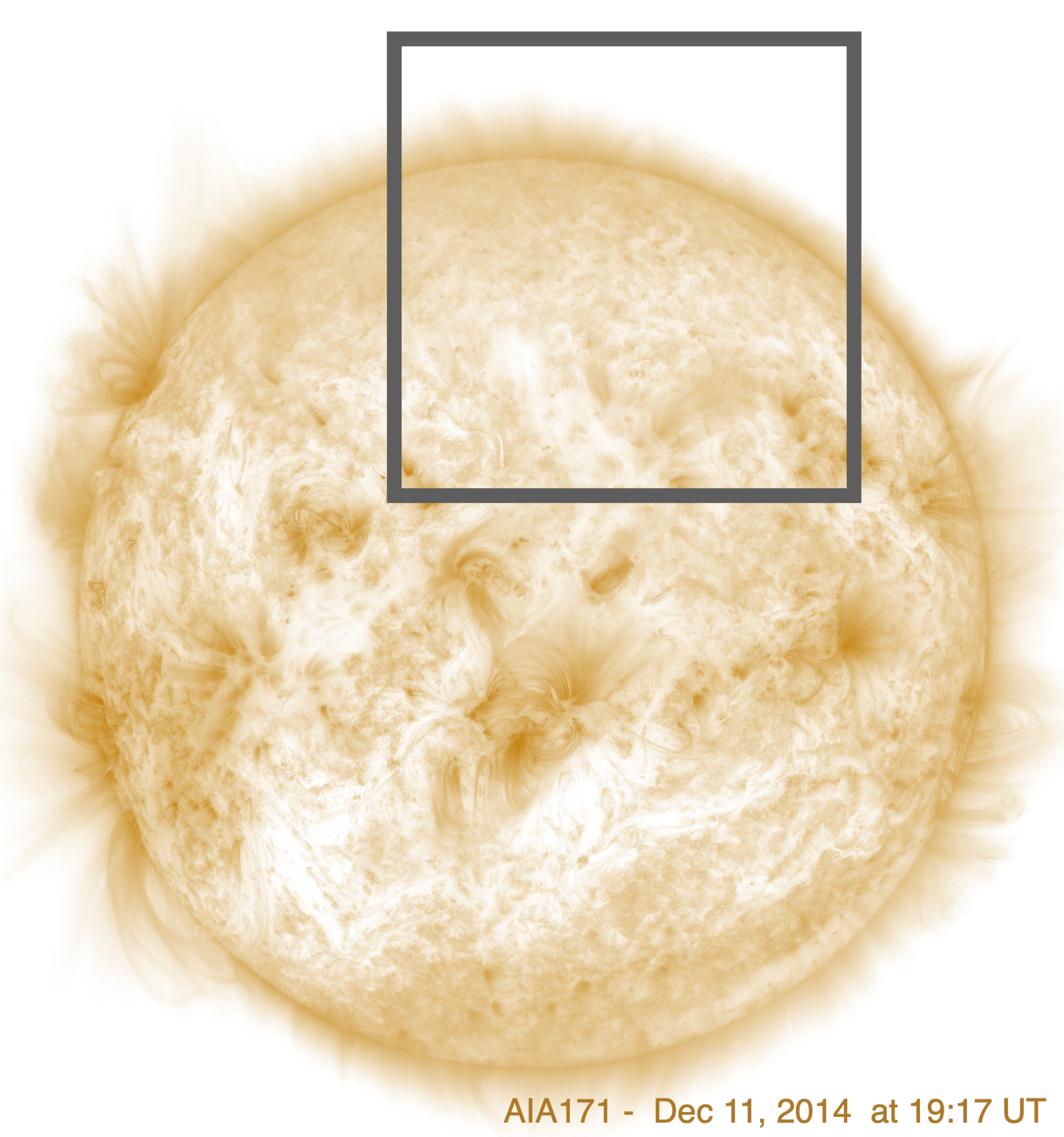}
\caption{FOXSI-2 quiet Sun target, at the north solar pole, observed during the FOXSI-2 rocket flight. The background image is the AIA solar full disk in the 171 Angstrom filter. The black square represents a sample of FOXSI's detectors FOV. The payload pointed to this target for a total of 92.7 seconds on December 11, 2014 (from 19:17:13.5 UTC to 19:18:46.2 UTC). The last 24.2s of this time were  used to measure background via shutters placed in front of the detectors.}
\label{Fig:quietsun_00}
\end{figure}

The FOXSI sounding rocket program is a mission to develop and test grazing-incidence HXR optics for solar observations. FOXSI uses a set of 7 Wolter-I figured grazing-incidence X-ray telescope modules to perform imaging spectroscopy of solar HXRs from 5-20 keV. The optics focal length, limited by the size of the sounding rocket payload (2 meters), sets the 20 keV energy upper limit.
The parameters of the optics, such as diameters and focal length, were set to suit the payload of a Terrier-Black-Brant sounding rocket. These optics were produced at NASA Marshall Space Flight Center applying a low-cost electroformed nickel alloy replication process, whereby nickel mirrors are electro-deposited onto super-polished mandrels \citep{ramsey2005replicated}. For increased effective area, shells of various radii are co-axially nested together into modules of 7 or 10 mirrors. The averaged resolution of the integrated modules was measured in the laboratory to be 4.3$\pm$0.6 arcsec (full width at half maximum, FWHM) and 27$\pm$1.7 arcsec (half-power diameter, HPD) for an on-axis source. Constrained to the Si detector square area, the field of view (FOV) is 16 $\times$ 16 arcmin$^2$.
A number of papers in the literature provide additional details about the FOXSI rocket experiment. \citet{krucker2013focusing,krucker2014first} describe the original payload and first scientific results of the mission. \citet{glesener2016foxsi} provide an overview of the first two flights of the experiment. \citet{christe2016foxsi} describe major updates made for the second flight as well as details on the mirror shell prescription. \citet{musset2019ghost}, \citet{athiray2017calibration} and \citet{furukawa2019development} describe the hardware upgrades for the 3rd flight of the sounding rockets. This paper is a continuation of the work described in \citet{buitrago2017methods} which describes implementations to reduce singly-reflected X-rays. Outstanding scientific results based on FOXSI’s observations are reported by \citet{ishikawa2017detection,athiray2020foxsi,Vievering2021ApJ}. \citet{buitrago2021foxsi} describe adaptation of the payload for a fourth rocket flight intended to observe a medium/large size solar flare in 2024. The GitHub repository \url{https://github.com/foxsi/foxsi-science} contains complete instructions to access and process FOXSI data collected during the first three rocket flights.

\section{Quiet Sun pointing with FOXSI-2}

The FOXSI rocket experiment has successfully flown three times from the White Sands Missile Range. The second launch (FOXSI-2) launched on December 11, 2014, at 19:11:00 UTC and targeted the Sun for 6 minutes and 40.8 seconds starting at 19:12:42 UTC. FOXSI-2's FOV was limited to about a quarter of the solar disk (see Figure \ref{Fig:quietsun_00}). We targeted five portions of the solar disk during the observation time to maximize science outcomes. The detailed list of FOXSI-2 targets is given in \cite{athiray2020foxsi}, and \cite{Vievering2021ApJ}. One of the targets covered a portion of the quiet Sun at the solar North pole for a total of 92.7 seconds (see Figure \ref{Fig:quietsun_00}). We will refer to this quiet Sun target as target \uppercase{I}, following the terminology coined by \cite{Vievering2021ApJ}. The dark grey box in Figure \ref{Fig:quietsun_00} is the FOV for one of the silicon detectors in FOXSI-2. All other silicon detectors in the payload had the same FOV size but were clocked in a set of different angles with respect to the one shown in Figure \ref{Fig:quietsun_00}.

For the last 24.2 seconds, pointing at target \uppercase{I}, we remotely activated an attenuator wheel that placed thick aluminum disks on top of the detectors for background measurements. Later, in section \ref{sec:ONOFF_Method}, we will use such background measurements to assess the existence, or not, of a source of HXRs of solar origin in target \uppercase{I}.

\section{Quiet Sun observation with FOXSI-3}
\label{sec:foxsi-3}

The FOXSI-3 rocket campaign took place at the White Sands Missile Range. The rocket launched on September 7, 2018, at 17:21 UT and observed the Sun from 17:22:44.6 UT until 17:29:14.1 UT, for a total of 6 minutes and 29.5 seconds. The payload contained seven Wolter-I optics modules paired with semiconductor detectors. See \cite{musset2019ghost} for details of the payload. Four detectors had silicon strip sensors; two other detectors worked with finer pitch CdTe strip sensors (60 $\mu$m instead of the 75 $\mu$m used in the silicon sensors). These six detectors were optimized for observations in the 4-20 keV energy range. The seventh detector (PhoEnIX) was a 2048$\times$2048 pixel CMOS sensor designed for soft X-ray observations (0.5 - 5.0 keV), see \cite{ishikawa2018high} for details.\\

\begin{figure}
\centering
\includegraphics[width=\hsize]{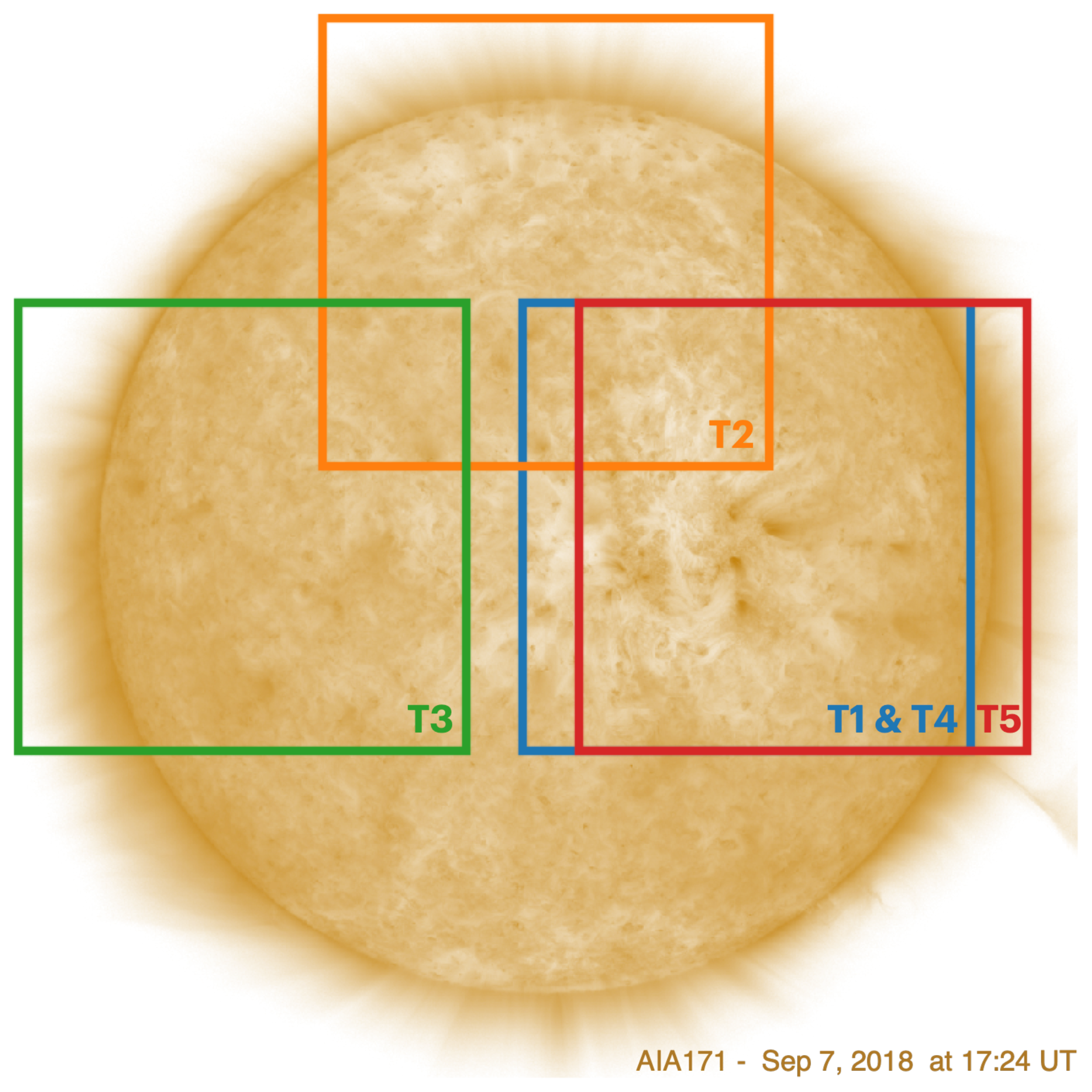}
\caption{The background image is the SDO/AIA 171\AA\,solar entire disk at the time of the FOXSI-3 observations (Sep 7, 2018, at 17:24 UT). Solar activity was very low at the time of the FOXSI-3 rocket launch. A very aged non-flaring active region was located in the western hemisphere. The colored squares represent the approximate FOV of a silicon detector and the targets during the FOXSI-3 observations. We highlight that these boxes are approximate FOVs because every detector is clocked differently.}
\label{Fig:FOXSI3}
\end{figure}

One of the primary goals for the FOXSI-3 rocket campaign was to place a more stringent HXR upper limit of the quiet Sun than previously reported. The launch of FOXSI-3 happened during a time of extremely low activity in the Sun. Figure \ref{Fig:FOXSI3} depicts an SDO/AIA solar image in 171 \AA\,at the time of the FOXSI-3 launch. A non-flaring aged active region can be identified in the western solar hemisphere. This active region was no longer intense enough to be labeled and included in the NOAA catalog. However, that active region existed for several solar rotations and was previously cataloged as NOAA AR12713 last time it was sufficiently active (June 2018). Figure \ref{Fig:FOXSI3} also shows a coronal hole at the solar north pole and sparse small EUV brightenings outside the aged active region.\\

FOXSI-3 pointed to the Sun and recorded data during 367.3 seconds total. FOXSI-3 targeted the aged active region for 128.2 seconds (blue T1 in Figure \ref{Fig:FOXSI3}), the north pole for 24.0 seconds (orange T2 in Figure \ref{Fig:FOXSI3}), the eastern quiet Sun limb for 144.6 seconds (green T3 in Figure \ref{Fig:FOXSI3}), and returned to the aged active region for 26.3 seconds (blue T4 in Figure \ref{Fig:FOXSI3}). The observations concluded with a 2 arcminutes shift towards the western limb where FOXSI-3 stayed for the remaining 44.2 seconds (red T5 in Figure \ref{Fig:FOXSI3}). Further details of the rocket campaign, and the upgrades in the payload, can be found in \cite{musset2019ghost}.\\

For the study using FOXSI-3 data presented here, we limited our analysis only to observations of three silicon detectors flown in the rocket. The fourth silicon detector included in the payload (as well as the two CdTe detectors) presented relatively high electronic noise during the flight, making them unsuitable for low counts analysis. Due to the low solar activity, HXRs recorded by FOXSI-3 were very sparse. The top part of Table \ref{tab:foxsi3observations} summarizes the total number of events observed with each of the three silicon detectors (D102, D105, and D106) in the 5-10 keV energy range, for every FOXSI-3 target (T1, T2, T3, T4, and T5).

Because of technical difficulties, we did not activated attenuators for the FOXSI-3 flight. The consequences of not having attenuators for some fraction of the observation time are that we do not have in-flight background measurements. 

\begin{figure*}
\includegraphics[width=\hsize]{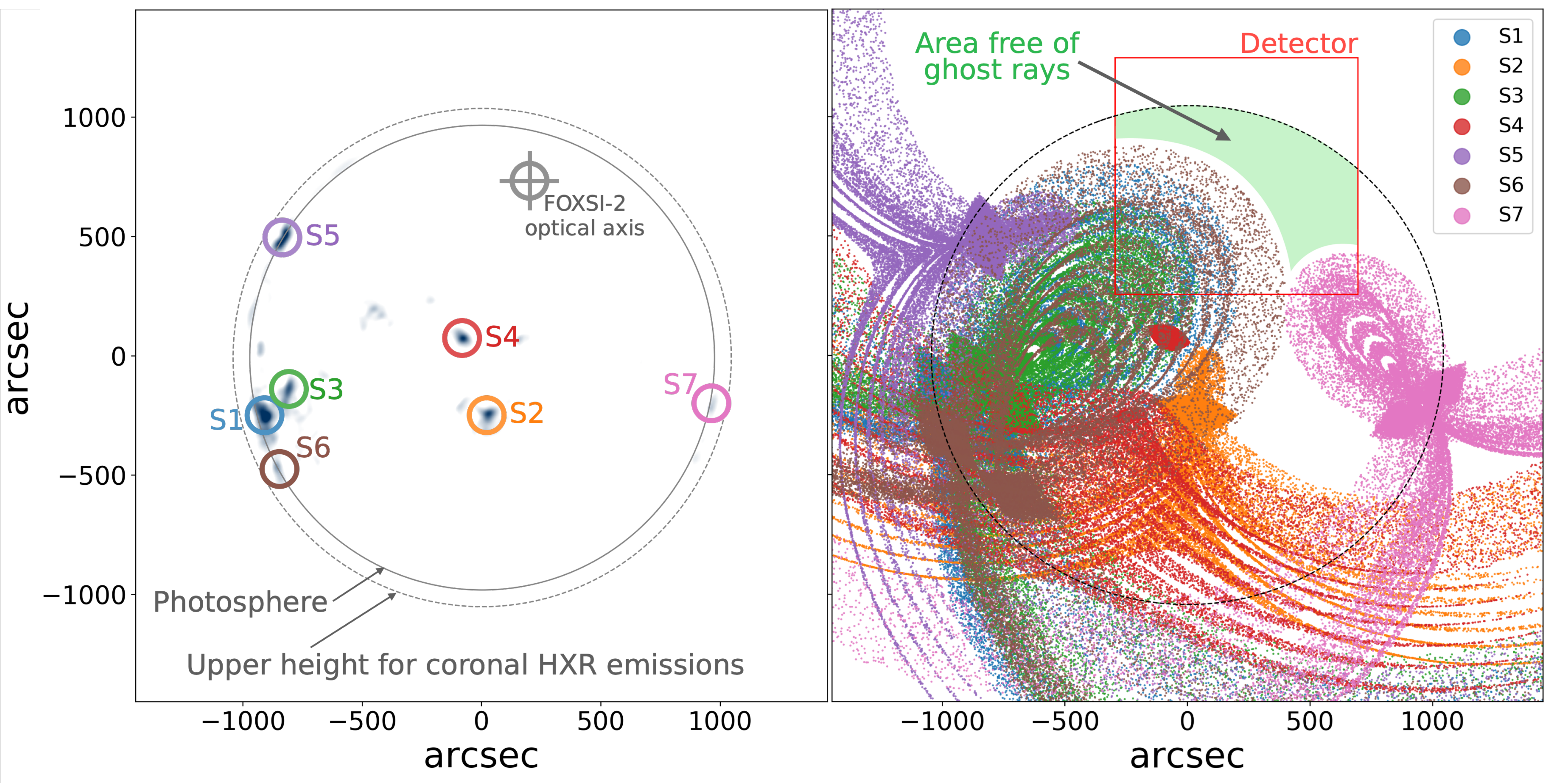}
\caption{\textit{Left}: Full disk FeXVIII map constructed from the 94, 171, and 211 AIA/SDO maps following \cite{del2013multi}. We identify seven intense, hot localized sources. We mark in grey the center of target \uppercase{I} at [200'', 750''] (FOXSI-2 optical axis). The solid black circle represents the photosphere. The dashed black circle sets the upper radius limit above which quiet Sun HXRs are not expected from (50 Mm above the photosphere). At this height, the ambient electron density gets lowered by more than four orders of magnitude compared to the photosphere, i.e., the HXR bremsstrahlung emission also gets substantially reduced \citep[see, e.g.,][]{aschwanden2006physics}. A few structures that are not circled, but seem as bright as others like S7, are ignored. The reason is that because of their short off-axis distances, their ghost rays are negligible, as is the case for sources S2, S4, and S5. \textit{Right}: Simulated ghost rays generated by the five intense sources when pointing to target \uppercase{I}. Each of the five source rays is color-coded according to the labels in the figure. The big black dashed circle represents the upper limit radius for coronal HXRs. The area in solid green sets the limit we chose as a region mostly free of ghost rays inside the solar disk. The red box is one of the silicon detector FOV. The other silicon detectors in FOXSI-2 had the same FOV size, but were clocked with respect to the one shown in this figure.}
\label{Fig:quietsun_01}
\end{figure*}

\section{Ghost ray treatment}
\label{sec:ghostrays}

FOXSI uses Wolter-I-figured grazing incidence X-ray telescopes to focus on solar X-rays. The Wolter-I geometry consists of two grazing-incidence mirror segments, a paraboloid primary mirror followed by a hyperboloid secondary reflector, referred to as mirror shells. On-axis rays that reflect on both mirrors are focused into an image on the focal plane. However, there is a possibility that rays from off-axis sources may reflect only on a single mirror shell and reach the focal plane. These single-reflecting rays are broadly referred to as stray light or ghost rays (see the right panel in Figure \ref{Fig:quietsun_01}). A full description of ghost rays for FOXSI, and strategies to minimize them (honeycomb structures used in FOXSI-3 for example), can be found in \cite{buitrago2017methods,musset2019ghost,buitrago2020use}.

To assess ghost rays polluting FOXSI's FOV, we need to know every identifiable, off-axis, intense, HXR source at the time of observing target \uppercase{I}. We constructed an AIA Fe XVIII map for December 11, 2014, at 19:17 UTC (target \uppercase{I} start time), using the method given by \cite{del2013multi}. This map, displayed as the background image of the right panel on Figure \ref{Fig:quietsun_01}, shows the hottest components of the coronal plasma at the time of our observations, i.e., potential sources of HXRs. From this Fe XVIII map, seven compact off-axis kernels are easily identifiable as potential sources of ghost rays. We used a FOXSI customized ray-tracing simulation \citep[see ][]{buitrago2020use} to assess the ghost ray effect that each of the seven compact sources has over the FOXSI-2 FOV. We show the results of such assessment in the right panel of Figure \ref{Fig:quietsun_01}. Although ghost rays impinge on a significant region within the detectors, they are constrained to an identifiable zone of the FOV. Our goal is not to characterize the intensity of ghost rays. Instead, we are trying to determine locations where ghost rays could reasonably be nonzero to exclude them from our analysis. Taking advantage of the ghost ray confinement, we can mask out ghost rays and define an area within the detector reasonably free of ghost ray light. Since our goal is to evaluate the solar origin of events FOXSI-2 observed, we selected a region, colored in green in the right panel of Figure \ref{Fig:quietsun_01}, from which coronal HXRs could originate. We study the events observed with FOXSI-2 in that region during the 64.5 seconds the rocket pointed to target \uppercase{I}, before activating the attenuators. Additionally, we use the last 24.2 seconds of target \uppercase{I} pointing as a background measurement.

\section{Statistical issue: assessing a weak source mixed with background data}

Typically, in high-energy astrophysics and physics, experiments measuring discrete sets of events (counts) may contain multiple signals (source(s) of interest mixed with background(s)). It is common practice to take additional auxiliary measurements to assess the background(s) by setting the experiment in a configuration believed to be free of the source(s) of interest. In these sets of measurements, the goal is usually to establish an actual count rate for the source(s) of interest. For reasonably large numbers of counts, many straightforward statistical background subtraction techniques are suitable to determine the existence of genuine sources \citep[see, e.g.,][]{mcivor2000background,piccardi2004background,benezeth2010comparative}. For faint sources and backgrounds, the measured counts are so few that usual Gaussian techniques based on normal distributions do not hold. Instead, Poisson and Binomial distributions appropriately describe low count statistics. \citeauthor{li1983analysis} in \citeyear{li1983analysis} published a first thorough review of this source and background low count statistics problem. Such a problem is today known as the ON/OFF problem or the Li-Ma problem. Although \citet{li1983analysis} proposed their statistical method originally in the context of gamma-ray astronomy, its generality is so wide that it can be directly applied to other fields in physics and astronomy. Particularly, the Li-Ma problem suits the FOXSI observations considered in this paper.

\section{ON/OFF Li-Ma analysis}
\label{sec:ONOFF_Method}

It is known that $N_{on}$ integer events measured by a counting experiment during a specific period of time follow the Poisson distribution \citep[see, e.g., ][]{li1983analysis,gehrels1986confidence,knoetig2014signal,casadei2014objective}:

\begin{equation}
    P(N_{on}|\lambda) = \frac{\lambda^{N_{on}}}{N_{on}!}e^{-\lambda},
\end{equation}

where $\lambda$ is the non-negative real number of expected events, a.k.a Poisson parameter. In the most simple ON/OFF problem, the Li-Ma problem, the $N_{on}$ measured counts are supposed to result from $s$ expected counts coming from a signal of interest overlaid with $b$ expected counts from a background. The ON/OFF Li-Ma framework assumes that $s$ and $b$ are independent Poisson variables, i.e., the sum $\lambda=s+b$ should follow a Poisson distribution with

\begin{equation}
    P(N_{on}|s,b) = \frac{(s+b)^{N_{on}}}{N_{on}!}e^{-(s+b)}.
\end{equation}

If $N_{off}$ represents the number of background counts measured when the experiment was set in an signal-off configuration, the distribution of such background counts is also a Poisson distribution with

\begin{equation}
    P(N_{off}|b') = \frac{b'^{N_{off}}}{N_{off}!}e^{-b'}.
\end{equation}

In general, the observation times for the on and off experiment configurations, $T_{on}$ and $T_{off}$, are not the same. To account for this difference, and others related to the details of the experiment setup (sensitive area $A$, detector livetimes $lt$, observed solid angle $\Omega$, etc.), a parameter $\alpha$ is introduced in the ON/OFF problem framework defined as  

\begin{equation}\label{eq:alpha}
    \alpha=\frac{T_{on}\cdot A_{on}\cdot lt_{on}\cdot \Omega_{on}\cdot ...}{T_{off}\cdot A_{off}\cdot lt_{off}\cdot \Omega_{off}\cdot ...},
\end{equation}

assumed to have negligible uncertainty \citep[see, e.g.,][]{berge2007background}. The expected counts from the background alone in the on- and off-signal of interest experiment setup ($b$ and $b'$ respectively) relate via $b=\alpha \,b'$.\\

Originally \citet{li1983analysis} proposed to assess the significance of a weak signal mixed with a background by use of a hypothesis test \citep[e.g., ][]{wilks1962mathematical,eadie1971statistical,gregory2005bayesian}. Later, \citet{knoetig2014signal} and \citet{casadei2014objective} developed such hypotheses test methods further, proposing objective Bayesian solutions for the ON/OFF Li-Ma problem using the three measurable quantities $N_{on}$, $N_{off}$, and $\alpha$.

For the hypothesis test method, $s$ and $b$ are the unknown parameters, and the \textit{null hypothesis} ($H_0$) is that $s\equiv 0$, i.e., the only signal is the background. The alternative hypothesis ($H_1$) considers $s > 0$. The conditional probability of $H_0$, $P(H_0|N_{on},N_{off},\alpha)$, is expressible in terms of Bayes' theorem \citep[e.g., ][]{knoetig2014signal},

\begin{equation}
    P(H_0|N_{on},N_{off},\alpha) = \frac{P(N_{on},N_{off}|H_0,\alpha)\,P_0(H_0)}{P(N_{on},N_{off}|\alpha)}.
\end{equation}

Here, $P(N_{on},N_{off}|H_0,\alpha)$ represents the probability of measuring $N_{on}$ and $N_{off}$, given a scenario where $H_0$ is true. $P(N_{on},N_{off}|\alpha)$ is a normalization probability, and $P_0(H_0)$ is the \textit{prior} probability for $H_0$.\\

There is a discussion among different authors regarding the effectiveness of different priors for the ON/OFF Li-Ma problem \citep[e.g., ][]{berger2001objective,casadei2014objective,nosek2016bayesian}. \citet{nosek2016bayesian} thoroughly analyzed the effect that three well-known priors (scale-invariant, uniform, and Jeffreys) have on the ON/OFF Li-Ma method when applied to weak signals. \citet{nosek2016bayesian} concluded that Bayesian inferences using Jeffreys' prior distributions are generally a safe compromise compared to the other priors they examined (scale invariant and uniform prior, for instance). Although Jeffreys' prior distributions require more complicated calculations based on integral expressions, it leads to reasonable limits of the source existence for close to zero observed counts. \citet{knoetig2014signal} implemented Jeffreys' prior and found an analytical solution to the ON/OFF Li-Ma problem in terms of special integral functions (Gamma and hypergeometric). The inputs of \citeauthor{knoetig2014signal}'s analytical solution are $N_{on}$, $N_{off}$, and $\alpha$. The outcomes are $P(H_0|N_{on},N_{off},\alpha)$, the Bayesian significance, $S_b=\sqrt{2}\,\text{erf}^{-1}[1-P(H_0|N_{on},N_{off},\alpha)]$, and a signal upper limit $\lambda_\sigma$, with an uncertainty of $\sigma$ (See the details of the general analytical solution in section 3.4 of \citet{knoetig2014signal}). \citeauthor{knoetig2014signal}'s solution is the one we implement here to analyse FOXSI-2 and -3 observations.

\section{ON/OFF Li-Ma analysis for FOXSI-2}

In the search for HXRs of quiet Sun origin, we applied Knoetig's solution of the ON/OFF Li-Ma problem to the observations of target \uppercase{I} in FOXSI-2. The first step is to set the off- and on-signal configurations. The off-signal observations occurred during the last 24.2 seconds of pointing to target \uppercase{I}, after the attenuators were activated, i.e., blocking the solar flux from reaching the experiment detectors. The sensitive area was the whole detector for the off-signal configuration, i.e., 16 $\times$ 16 arcmin$^2$. The on-signal observations consist of the counts recorded by a detector (during the 68.5 seconds of no-attenuators) within the green area described in the right panel of Figure \ref{Fig:quietsun_01}.\\

\begin{figure}
\centering
\includegraphics[width=\hsize]{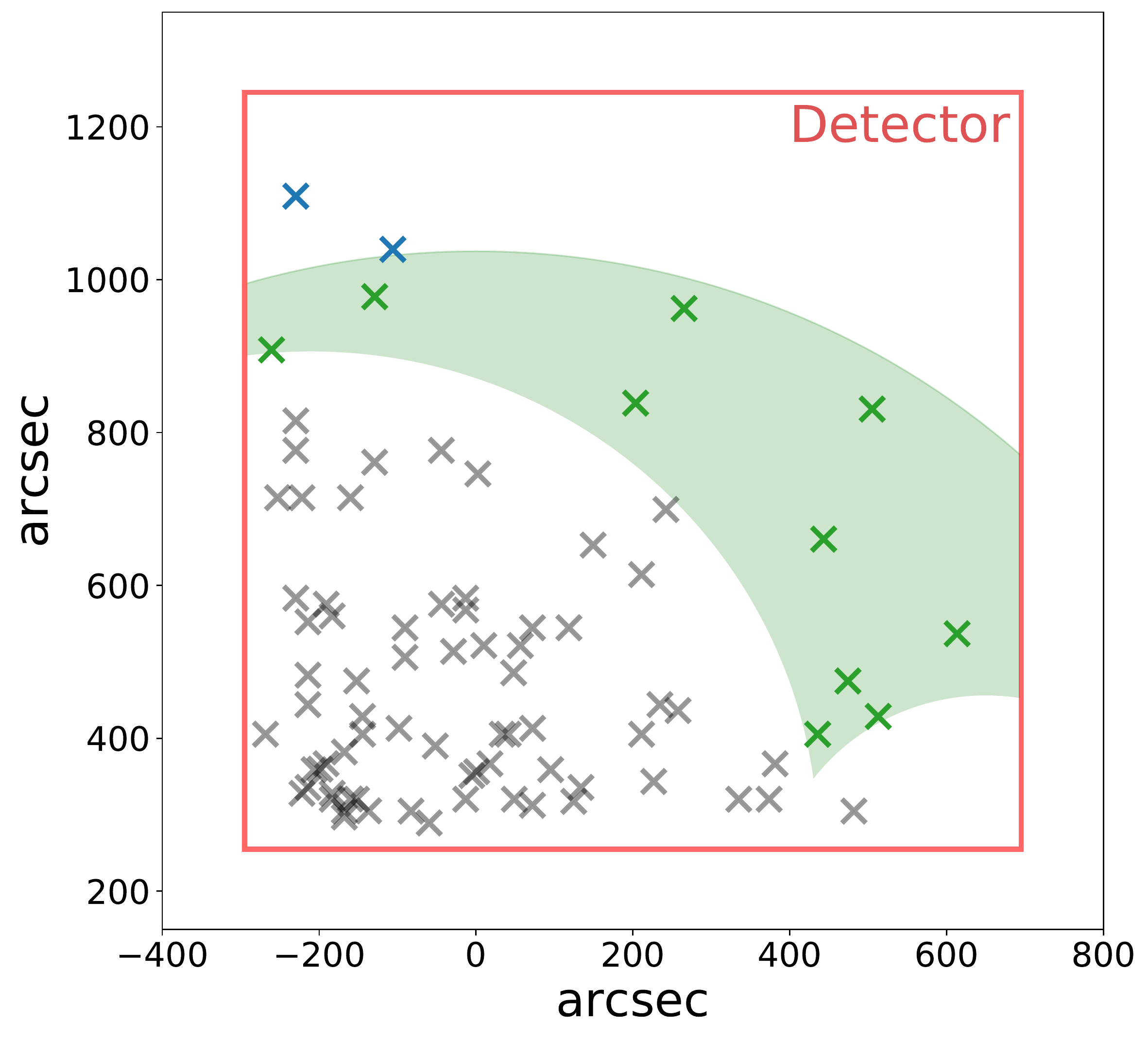}
\caption{Counts registered by one of the FOXSI-2 silicon detectors when pointing to target \uppercase{I}. The area in solid green sets the limit we chose as a region mostly free of ghost rays inside the solar disk (see Figure \ref{Fig:quietsun_01} for reference). The red square shows one detector FOV. All Xs in the plot are events recorded for one of the detectors. The dots are black if they are classified as ghost rays, green if the events fall within the solar region free of ghost rays, and blue if they are outside the solar disk.}
\label{Fig:quietsun_02}
\end{figure}

FOXSI-2 flew seven optics/detector assemblies. However, here we use only the most reliable four detectors to apply the ON/OFF Li-Ma analysis (D101, 104, D105, and D108 hereafter). Of the remaining three detectors, two were a bit noisy for weak sources studies, and one was placed on a location in the payload with no attenuator (i.e., with no background measurement).
All four detectors we use in this study had a silicon sensor and were positioned behind a 7-mirror optics module. Figure \ref{Fig:quietsun_02} shows the counts observed by one of these detectors (D105) during the on-signal configured observation. The ten green dots in Figure \ref{Fig:quietsun_02} constitute $N_{on}$. $N_{off}$ for that same detector is four counts. We calculated $\alpha$ as the ratio of observation time (corrected by the detector livetimes) and the observed areas between the on- and off-configuration. For D105, $\alpha = 0.86$. These values, and the ones for the other three analyzed detectors, are summarized in Table \ref{tab:foxsi2}. Table \ref{tab:foxsi2} also displays the outcomes of \citeauthor{knoetig2014signal}'s method for the four detectors: The probability of $H_0$ ($P(H_0|N_{on},N_{off},\alpha)$), the significance $S_b$, and the upper limit with a $2\sigma$ (97.72\%) confidence.\\

\begin{table}
\centering
\resizebox{\columnwidth}{!}{%
\begin{tabular}{l|c|c|c|c|c|}
\cline{2-6}
& \multicolumn{1}{c|}{{\color[HTML]{EA761D} D101}} & \multicolumn{1}{c|}{{\color[HTML]{359D1A} D104}} & \multicolumn{1}{c|}{{\color[HTML]{C32D25} D105}} &
\multicolumn{1}{c|}{{\color[HTML]{3D85CC} D108}} &
\multicolumn{1}{c|}{{\color[HTML]{5E5F5F} All four Det }}\\ \hline
\multicolumn{1}{|c|}{$N_{ON}$ ($\Delta t_{ON} = 68.5$ s)} & {\color[HTML]{EA761D} 13} & {\color[HTML]{359D1A} 9} & {\color[HTML]{C32D25} 10} & {\color[HTML]{3D85CC} 10} & {\color[HTML]{5E5F5F} 36 }\\ \hline
\multicolumn{1}{|c|}{$N_{OFF}$ ($\Delta t_{OFF} = 24.2$ s)} & {\color[HTML]{EA761D} 3} & {\color[HTML]{359D1A} 1} & {\color[HTML]{C32D25} 1} & {\color[HTML]{3D85CC} 4} & {\color[HTML]{5E5F5F} 9 }\\ \hline
\multicolumn{1}{|c|}{$\alpha$} & {\color[HTML]{EA761D} 0.920} & {\color[HTML]{359D1A} 0.872} & {\color[HTML]{C32D25} 0.869} & {\color[HTML]{3D85CC} 0.859} & {\color[HTML]{5E5F5F} 0.887 } \\ \hline
\multicolumn{1}{|c|}{$P(H_0|N_{ON},N_{OFF},\alpha)$} & {\color[HTML]{EA761D} $8.3\times10^{-3}$} & {\color[HTML]{359D1A} $5.5\times10^{-3}$} & {\color[HTML]{C32D25} $2.8\times10^{-3}$} & {\color[HTML]{3D85CC} $6.2\times10^{-2}$} & {\color[HTML]{5E5F5F} $1.2\times10^{-5}$ } \\ \hline
\multicolumn{1}{|c|}{$S_b$} & {\color[HTML]{EA761D} 2.64} & {\color[HTML]{359D1A} 2.78} & {\color[HTML]{C32D25} 2.99} & {\color[HTML]{3D85CC} 1.86} & {\color[HTML]{5E5F5F} 4.4 } \\ \hline
\multicolumn{1}{|c|}{$\lambda_{2\sigma}$} & {\color[HTML]{EA761D} 19.09} & {\color[HTML]{359D1A} 15.52} & {\color[HTML]{C32D25} 16.83} & {\color[HTML]{3D85CC} 14.73} & {\color[HTML]{5E5F5F} 10.5 } \\ \hline
\end{tabular}%
}
\caption{Input (three first rows) and output (three last rows) parameters of the ON/OFF Li-Ma analysis applied to four of the FOXSI-2 silicon detectors (first four columns). The right most column has the ON/OFF Li-Ma parameters for the case in which we combine data of all four detectors. $N_{ON}$ are the counts recorded during the ON configuration of target \uppercase{I}/FOXSI-2 observations (green area of Figure \ref{Fig:quietsun_01}). $N_{OFF}$ are the counts for the background observations. $\alpha$ is estimated following equation \ref{eq:alpha} (for individual detectors) and equation \ref{eq:alphaT} (for the case of all four detectors combined). $P(H_0|N_{ON},N_{OFF})$ are the probabilities that the null hypothesis ($H_0$) is true given the particular values of $N_{ON}$, $N_{OFF}$ and $\alpha$. $S_b$ is the Bayesian significance for the existence of a hypothetical signal $s$ on top of the background during the ON-configured observations. $\lambda_{2\sigma}$ is the upper limit (in counts) with a $2\sigma$ precision for the flux of such a hypothetical source $s$.}
\label{tab:foxsi2}
\end{table}

\begin{figure}
\centering
\includegraphics[width=\hsize]{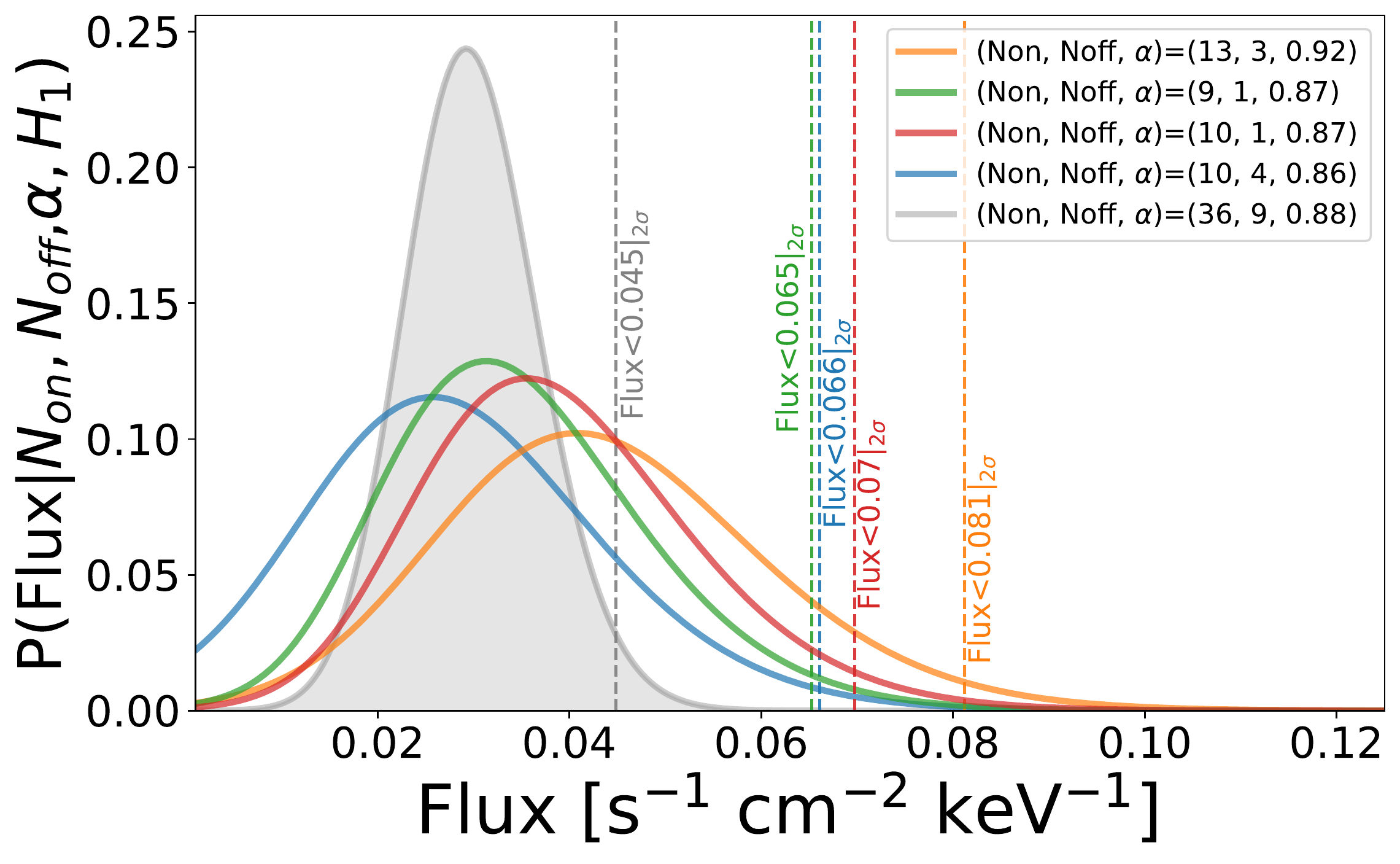}
\caption{Source flux distribution functions for four of the silicon detectors in FOXSI-2. The lines are colored to match Table \ref{tab:foxsi2}. All four colored distribution functions show a similar profile. The slight differences among the lines can be attributed to low count statistical effects. The dashed vertical lines are the upper limits with a $2\sigma$ certainty for each of the four distribution functions. The gray-filled curve is the normalized distribution function putting together the observations of all four detectors (accounting for each detector livetime and average optics effective area). The corresponding upper limit (gray dashed line) corresponds to a HXR solar flux of $>$0.045 s$^{-1}$ cm$^{-2}$ keV$^{-1}$. The maximum value of the gray distribution function lies at $\sim$0.029 s$^{-1}$ cm$^{-2}$ keV$^{-1}$.}
\label{Fig:f2_probabilities}
\end{figure}

\citeauthor{knoetig2014signal}'s method produces the distribution function for $s$, i.e., the probability  $P(s|N_{on},N_{off},\alpha,H_1)$ as a function of the expected signal of interest counts. We can express such a distribution function in terms of the HXR flux from the whole Sun by using the conversion

\begin{equation}\label{eq:flux}
    Flux = \frac{s}{\Delta t\,\Delta E\,\Delta A}\frac{\Delta\Omega_{\odot}}{\Delta \Omega_{on}},
\end{equation}

where $\Delta t$ is the observation time (corrected by the detector livetime), $\Delta E$ is the observed energy bandwidth, and $\Delta A$ is the optics effective area averaged over the energy bands considered (5-10 keV). We additionally correct by the scale ratio of solid angles ($\frac{\Delta\Omega_{\odot}}{\Delta \Omega_{on}}$) to get an estimate of the flux over the whole observable solar corona. Using the conversion in expression \ref{eq:flux} we plot the probability distribution function for each of the four detectors in Figure \ref{Fig:f2_probabilities}. Additionally, Figure \ref{Fig:f2_probabilities} displays vertical dashed lines indicating the upper limits with a $2\sigma$ certainty for distribution functions of each of the four detectors. Regardless of the very low count statistics, all four colored distribution functions in Figure \ref{Fig:f2_probabilities} exhibit similar behaviours, with maximum probabilities around $\sim$0.03 s$^{-1}$ cm$^{-2}$ keV$^{-1}$ and comparable upper limits.

The gray filled curve in Figure \ref{Fig:f2_probabilities} corresponds to the normalized source distribution function, versus the HXR solar flux, using data from all four detectors put together (accounting for the respective livetimes and effective areas of each detector/optics ensemble). To compute the ON/OFF Li-Ma analysis for the combination of the four detectors, $\alpha$ is transformed into $\alpha_{combined}$ defined as

\begin{equation}\label{eq:alphaT}
    \alpha_{combined} = \frac{\sum_d T_{on}\cdot A_{on}\cdot lt_{on}\cdot \Omega_{on}}{\sum_d T_{off}\cdot A_{off}\cdot lt_{off}\cdot \Omega_{off}},
\end{equation}

where $\sum_d$ is the sum over each of the four detector/optics sets. When performing the single ON/OFF Li-Ma analysis for the four detectors combined, the statistical significance of the measurements improves, $S_b$ = 4.4 (see Table \ref{tab:foxsi2}). In general, such a high significance suggests the detection of a signal. However, we abstain from claiming that that signal comes from the quiet Sun for this particular case. The reason is that although we characterized the most severe sources of ghost rays, there may still be stray light of other origins that we are not accounting for. Such possible additional stray light could come from a relatively slight misalignment among the optics module axes (that we believe to be under $\sim$1.5 arcminutes based on measurements performed before and after the rocket flight). The optics point spread function's wings could also be another source of extra faint stray light. Within the calibration resources available in a sounding rocket program, it is not possible to completely rule out the presence of ghost rays in our measurement area. Instead, we report an upper limit for a signal of solar origin. Our careful ghost ray treatment enables this to be a highly sensitive limit. From the analysis that uses data from all four detectors, we found an upper limit for the HXR quiet Sun flux of 0.045 s$^{-1}$ cm$^{-2}$ keV$^{-1}$ with a 97.72\% (2$\sigma$) certainty (dashed vertical grey line in Figure \ref{Fig:f2_probabilities}). If we want to be conservative with the claims in this work, we can instead choose the largest of the four limits shown in Figure \ref{Fig:f2_probabilities}. That conservative limit is 0.081 s$^{-1}$ cm$^{-2}$ keV$^{-1}$ (dashed vertical orange line in Figure \ref{Fig:f2_probabilities}).

\section{Statistical analysis of the FOXSI-3 quiet sun observations}
\label{sec:hola}

Because of the lack of in-flight background measurements during the FOXSI-3 observation, we used two alternatives to analyze the quiescent data collected during that flight. The first (\citeauthor{gehrels1986confidence}' method hereafter) sets upper limits assuming a source (or mix of weak sources) described with a Poisson distribution. For our second approach, we use background measurements taken in FOXSI-2 to set upper limits on the FOXSI-3 observations.

\vspace{.5cm}

\begin{table}[]
\centering
\resizebox{\columnwidth}{!}{%
\begin{tabular}{l|c|c|c|c|c|}
\cline{2-6}
& \multicolumn{1}{c|}{D102}  & \multicolumn{1}{c|}{D105}  & \multicolumn{1}{c|}{D106} & \multicolumn{1}{c|}{All three Det} & \multicolumn{1}{c|}{Obs. time [s]} \\\hline
\multicolumn{1}{|l|}{T1 counts}  & 3 & 0 & 3 & 6 & 128.2 \\ \cline{1-1}
\multicolumn{1}{|l|}{T2 counts}  & 0 & 1 & 1 & 2 & 24.0 \\ \cline{1-1}
\multicolumn{1}{|l|}{T3 counts}  & 1 & 1 & 0 & 2 & 144.6 \\ \cline{1-1}
\multicolumn{1}{|l|}{T4 counts}  & 1 & 0 & 1 & 2 & 26.3 \\ \cline{1-1}
\multicolumn{1}{|l|}{T5 counts}  & 0 & 0 & 0 & 0 & 44.2 \\ \hline
\multicolumn{1}{|l|}{All target counts} & \multicolumn{1}{c|}{5} & \multicolumn{1}{c|}{2} & \multicolumn{1}{c|}{5} & \multicolumn{1}{c|}{12} & 367.3 \\\hline
\end{tabular}
}
\caption{FOXSI-3 observation summary. Counts recorded in the 5-10 keV energy range with three silicon detectors (D102, D105, and D106). T1-T5 are the targets pointed during the rocket observations according to figure \ref{Fig:FOXSI3}. The second to the right most column shows the sum of the counts observed with the three detectors.}
\label{tab:foxsi3observations}
\end{table}

\subsection{\citeauthor{gehrels1986confidence}' method to set upper quiet Sun limits for FOXSI-3}

\cite{gehrels1986confidence} provided a set of upper limit tables for hypothetical signal rates as the source(s) of a small number of observed events. We used the total number of counts observed by each of the three silicon detectors, summarized in Table \ref{tab:foxsi3observations}, to set the upper limits in Table \ref{tab:foxsi3gehrels} for a $2\sigma$ confidence level. Table \ref{tab:foxsi3gehrels} also contains the upper limit flux ($F_{2\sigma}$) computed using a modified version of equation \ref{eq:flux} where we replace $s$ in the expression with the values of $\lambda_{2\sigma}$:

\begin{equation}\label{eq:fluxlimit}
    F_{2\sigma} = \frac{\lambda_{2\sigma}}{\Delta t\,\Delta E\,\Delta A}\frac{\Delta\Omega_{\odot}}{\Delta \Omega_{on}}.
\end{equation}

Despite the impossibility of doing background removal, applying \citeauthor{gehrels1986confidence}' method over the more than six minutes of observation time during FOXSI-3 gives us upper limits that are over two orders of magnitude lower than what we found for FOXSI-2. This is further evidence that the region identified as free of ghost rays may still contain background X-rays of solar or non-solar origin. Later, in section \ref{Discussion_and_conclusions} we discuss whether this difference may lie in an intrinsic correlation of the quiescent emission in HXRs with the phases of the solar cycle.

\begin{table}[]
\centering
\resizebox{\columnwidth}{!}{%
\begin{tabular}{l|c|c|c|c|}
\cline{2-5}
& \multicolumn{1}{c|}{D102}  & \multicolumn{1}{c|}{D105}  & \multicolumn{1}{c|}{D106} & \multicolumn{1}{c|}{All three Det} \\\hline
\multicolumn{1}{|c|}{$\lambda_{2\sigma}$} & \multicolumn{1}{c|}{11.8} & \multicolumn{1}{c|}{7.3} & \multicolumn{1}{c|}{11.8} & \multicolumn{1}{c|}{21.16}\\\hline
\multicolumn{1}{|c|}{$F_{2\sigma}$ [s$^{-1}$cm$^{-2}$keV$^{-1}$]} & \multicolumn{1}{c|}{2.4$\times10^{-3}$} &
\multicolumn{1}{c|}{1.3$\times10^{-3}$} & \multicolumn{1}{c|}{4.1$\times10^{-3}$} & 
\multicolumn{1}{c|}{6.0$\times10^{-4}$}\\\hline
\end{tabular}
}
\caption{FOXSI-3 upper limits with a $2\sigma$ confidence level ($\lambda_{2\sigma}$, in counts) evaluating the existence of a hypothetical signal present during the observations. These upper limits are directly extracted from the tables in \cite{gehrels1986confidence} for 5, 2, 5, and 12 counts respectively. $F_{2\sigma}$ are the  HXR solar fluxes estimated from the $\lambda_{2\sigma}$ values when computed with the instrument response.}
\label{tab:foxsi3gehrels}
\end{table}

\subsection{ON/OFF Li-Ma analysis on the FOXSI-3 observations}

\begin{table}[ht!]
\centering
\resizebox{\columnwidth}{!}{%
\begin{tabular}{|c|c|c|c|c|c|c|}
\hline
\multicolumn{7}{|c|}{D105}\\ \hline
$N_{ON}$ & $N_{OFF}$ & $\alpha$ & P($H_0$) & $S_b$   & $\lambda_{2\sigma}$  & $F_{2\sigma}$ [s$^{-1}$cm$^{-2}$keV$^{-1}$] \\\hline
2   & 1    & 16.1  & 0.89 & 0.14 & 5.43 & $9.6\times10^{-4}$ \\\hline
\end{tabular}
}
\caption{Summary table for the input and output parameters of the ON/OFF Li-Ma method applied exclusively to D105 using solar observations from FOXSI-3 and background measurements from FOXSI-2. $N_{ON}$ are the number of events observed by D105 during the entire 6.49 minutes of observation of FOXSI-3. $N_{OFF}$ are the counts register by D105 during the 24.2 seconds the attenuators were activated during FOXSI-2. $\alpha$ is calculated according to equation \ref{eq:alpha}. $P(H_0|N_{ON},N_{OFF})$ is the probability that the null hypothesis ($H_0$, for $N_{ON}$, $N_{OFF}$, and $\alpha$ given). $S_b$ is the Bayesian significance for the existence of an hypothetical quiet Sun signal $s$. $\lambda_{2\sigma}$ is the upper limit (in counts) with a $2\sigma$ confidence level for such a hypothetical source $s$. $F_{2\sigma}$ is the same upper limit but in units of s$^{-1}$cm$^{-2}$keV$^{-1}$.}
\label{tab:foxsi3}
\end{table}

\begin{figure}
\centering
\includegraphics[width=\hsize]{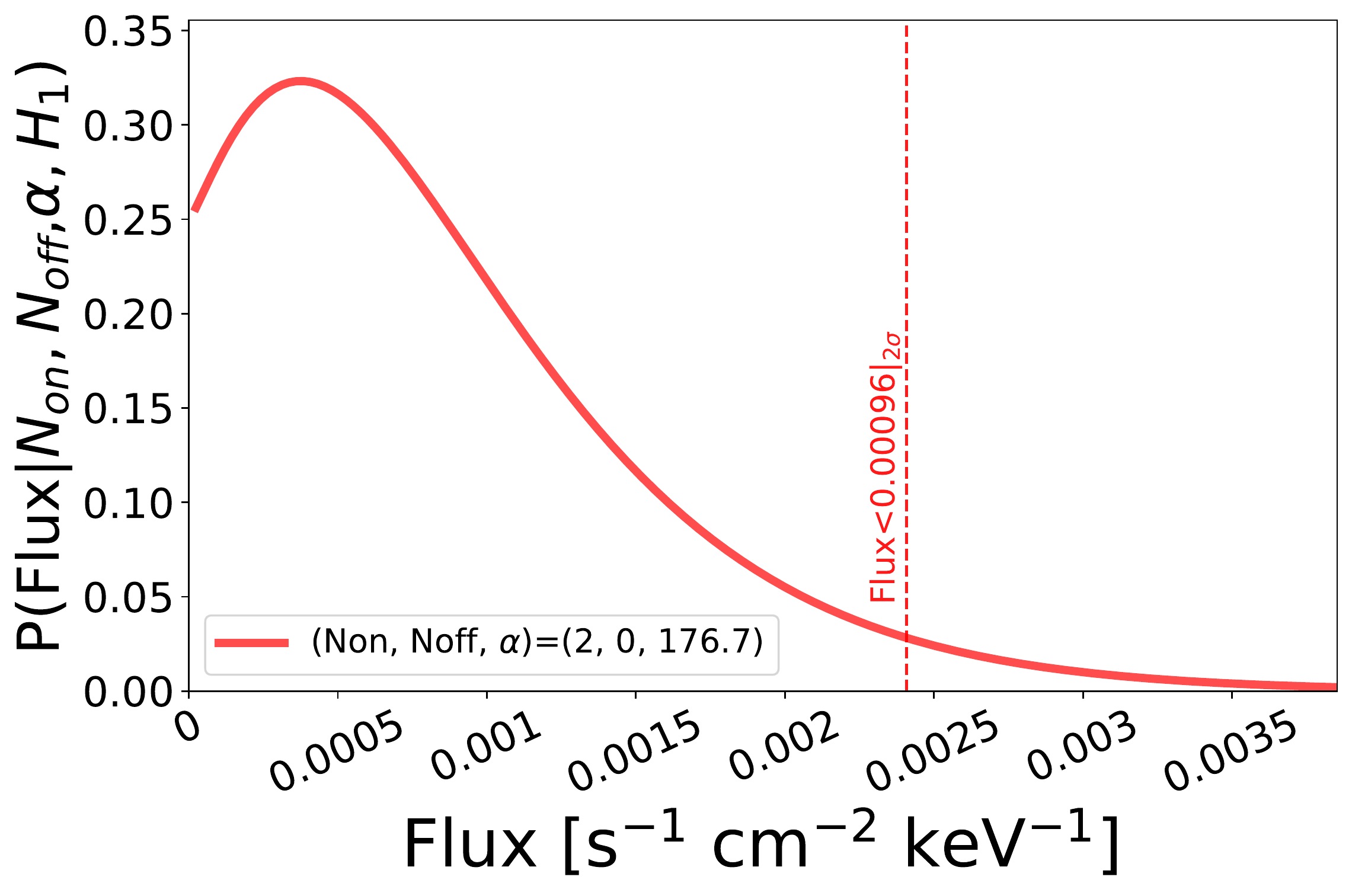}
\caption{Probability distribution function of a hypothetical HXR solar source as a function of its expected flux. This curve is constructed by implementing the ON/OFF Li-Ma method using FOXSI-3 observations and FOXSI-2 background measurements as the ON and OFF configurations, respectively. This figure corresponds to data registered by D105 which was the only silicon detector flown in FOXSI-2 and -3.}
\label{Fig:Det4DistFunction}
\end{figure}

To implement the ON/OFF Li-Ma method, it is critical to have an OFF configured observation, i.e., a background measurement. We identified D105 as the single silicon detector flown in both FOXSI-2 and FOXSI-3 rocket campaigns. The FOXSI-2 and FOXSI-3 instruments were launched using the same type of rocket, comparable trajectory parameters, and similar ambient conditions for the payload. We can argue that because of the similarities of the two flights, we can use D105 background measurements from FOXSI-2 to apply the ON/OFF Li-Ma method with D105 observations recorded during FOXSI-3. Table \ref{tab:foxsi3} summarizes the result of applying such an ON/OFF Li-Ma analysis. We highlight that the HXR solar flux with a $2\sigma$ confidence level obtained with this method, $F_{2\sigma}=9.6\times10^{-4}$ s$^{-1}$ cm$^{-2}$ keV$^{-1}$, is of the same order of magnitude as the one found using \citeauthor{gehrels1986confidence}' method (which does not assume a background). The fact that the FOXSI-3 observation time was over six times longer than that of FOXSI-2 causes $\lambda_{2\sigma}$ to be significantly reduced (in this case, around 70 times smaller). Figure \ref{Fig:Det4DistFunction} shows the distribution function for D105 according to the ON/OFF Li-Ma analysis. There are two remarkable traits in Figure \ref{Fig:Det4DistFunction}. i) The distribution peak is at zero, consistent with the 89\% probability that the null hypothesis is true. The null hypothesis demands no quiet Sun sources observed in the FOXSI-3 data, i.e., all counts being solely background during the ON configured measurements. ii) Consequently, the upper limits dramatically shift to lower values than those obtained for  FOXSI-2 (see Figure \ref{Fig:f2_probabilities}). There is a difference of over two orders of magnitude between the limits of FOXSI-2 and FOXSI-3. This difference, again, is the product of longer observation times, larger collecting areas (detectors free of ghost rays), and the fact that the whole Sun was quiet for the FOXSI-3 launch.\\

In sections \ref{sec:ghostrays} and \ref{sec:ONOFF_Method}, we presented a thorough analysis to identify regions free of ghost rays during the FOXSI-2 observations. For FOXSI-3, ghost rays are not a concern. The reasons lie in the fact that the Sun exhibited an extremely quiescent atmosphere for FOXSI-3 compared to FOXSI-2. During the FOXSI-3 observations, we scanned most of the solar disk (including the aged active region, which was the hottest part at the time) as depicted in Figure \ref{Fig:FOXSI3}. During those observations, we did not find a single discernible intense compact source in HXRs. FOXSI-3 only registered sparse data, as shown in Table \ref{tab:foxsi3observations}. According to \cite{buitrago2020use}, ghost-ray intensities are one order of magnitude fainter than their focused counterparts. Therefore, any ghost-ray background in FOXSI-3 would have been one order of magnitude lower than what we observed within the detector's field of views. Since for FOXSI-3, both ghost rays and on-axis photons would have the same origin (quiet Sun HXRs), the upper limits we report still hold.



\section{Comparing our upper limits with those previously reported}

\begin{figure}
\centering
\includegraphics[width=\hsize]{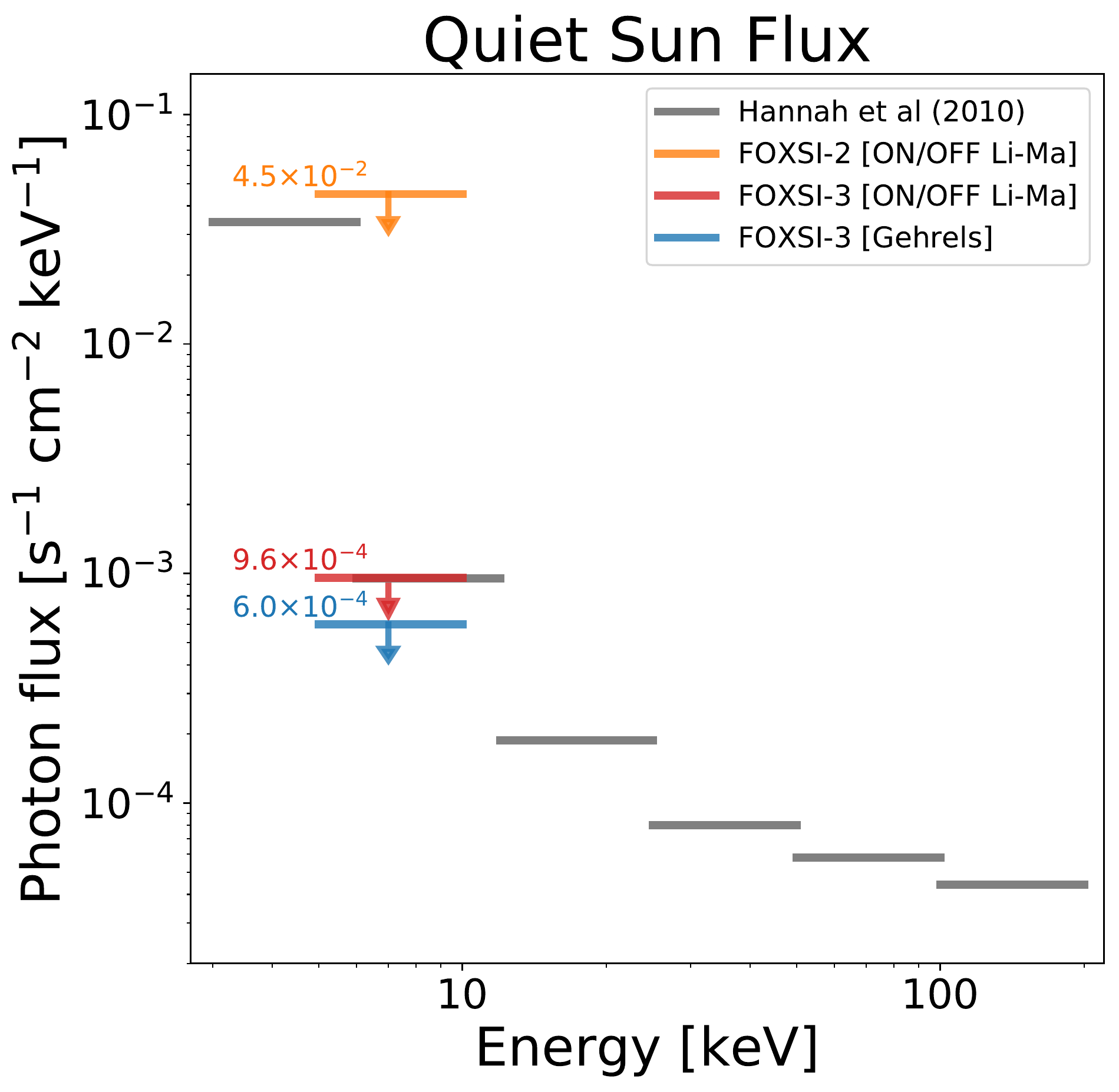}
\caption{Upper limits of the quiet Sun photon flux spectrum. The thresholds in gray are taken from \cite{hannah2010constraining}. They correspond to $2\sigma$ limits calculated based on the analysis of 11.9 days of solar off-pointing observations with RHESSI during solar quiescent conditions. We overlap three upper limits we found using FOXSI-2 and -3 data, all three in the 5-10 keV energy range. The limit in orange ($4.5\times10^{-2}$ s$^{-1}$cm$^{-2}$keV$^{-1}$) is calculated by implementing the ON/OFF Li-Ma method over an area free of ghost rays during $\sim 1$ minute of FOXSI-2 observations. The threshold in red ($9.6\times10^{-4}$ s$^{-1}$cm$^{-2}$keV$^{-1}$) corresponds to the upper limit obtained by combining FOXSI-3 measurements of only one detector that also had background measurements from the FOXSI-2 flight. This limit was also computed by implementing the ON/OFF Li-Ma method. The blue bar ($6.0\times10^{-4}$ s$^{-1}$cm$^{-2}$keV$^{-1}$) is the upper threshold estimated from the 6.49 minutes of observations with three FOXSI-3 silicon detectors combined. For this last threshold (blue), we used \cite{gehrels1986confidence} approach to set the upper expected rate of a hypothetical source of solar origin.}
\label{Fig:hannah}
\end{figure}

\cite{hannah2010constraining} used $\sim$12 days of off-pointing RHESSI data to estimate upper limits for HXR quiet-sun emission. Those $\sim$12 days of data correspond to observations during a period of solar cycle minimum. \cite{hannah2010constraining} reported upper thresholds for the photon flux as shown in the Figure \ref{Fig:hannah}. We overlap in the same Figure \ref{Fig:hannah} three upper limits from our analyses. In orange we plot the upper limit we obtained by combining FOXSI-2 data from four silicon detectors ($4.5\times10^{-2}$ s$^{-1}$cm$^{-2}$keV$^{-1}$). In red, we show the upper limit we calculate using the ON/OFF Li-Ma method applied to D105 observations during FOXSI-3, and background measurements from FOXSI-2 (that is $9.6\times10^{-4}$ s$^{-1}$cm$^{-2}$keV$^{-1}$). The blue upper limit in Figure \ref{Fig:hannah} ($6.0\times10^{-4}$ s$^{-1}$cm$^{-2}$keV$^{-1}$) is computed using \citeauthor{gehrels1986confidence}' method over data from three silicon detectors flown in FOXSI-3. For reference, \citeauthor{hannah2010constraining} upper limits for the 3-6 keV and 6-12 keV are $3.4\times10^{-2}$ s$^{-1}$cm$^{-2}$keV$^{-1}$ and $9.5\times10^{-4}$ s$^{-1}$cm$^{-2}$keV$^{-1}$, respectively. The FOXSI-2 and -3 limits found in this work are similar to with the deepest limits for solar HXR emission yet reported \citep{hannah2010constraining}.\\

\citet{hannah2010constraining} used the limits they found with RHESSI to constrain the parameter space of an isothermal model and thin-target emission models (power-law and kappa distributions) for the solar corona. \citet{hannah2010constraining} showed with their limits that it is unlikely for nanoflares with non-thermal effects to be involved in heating of the quiet corona. They concluded that such nanoflares would require a steep electron spectrum $E^{-\delta}$ with $\delta > 5$ extending to very low energies into the thermal energy range (<1 keV). Remarkably, the upper limits we found using FOXSI-3 data are in statistical agreement with those reported by \citet{hannah2010constraining}. Thus, all the conclusions asserted by \citet{hannah2010constraining} about the nature of nanoflares in the quiet Sun still hold. 

\section{Discussion and conclusions}
\label{Discussion_and_conclusions}

In this paper, we provided for the first time a quantitative limit of the quiet Sun HXR flux using data taken with the FOXSI sounding rocket instrument exclusively. We used data from FOXSI's second and third flights, corresponding to high and low solar cycle activity periods, respectively.

Because of the high solar activity during the FOXSI-2 flight, the Sun had several bright compact HXR sources distributed all over the disk. When located off-axis, such compact sources produced ghost rays extending partially into FOXSI's detector areas. We characterized the ghost rays impact on the instrument by implementing a ray-tracing simulation. Using those algorithms, we identified areas within the detectors mostly free of ghost rays. This approach allowed us for the first time to assess the flux in HXRs of a quiescent solar region during a time of substantially high solar activity. This is something that has never been possible to do before with solar HXR telescopes that use indirect imaging techniques, like RHESSI. We implemented a Bayesian analysis optimized for very low statistics (the Li-Ma ON/OFF method) to estimate an upper threshold of 0.045 s$^{-1}$cm$^{-2}$keV$^{-1}$ for the HXR flux (5-10 keV) within the identified quiet Sun area almost entirely free of ghost rays.

This paper is also the first science work that uses FOXSI-3 data. The Sun was at solar minimum when FOXSI-3 flew. Only a very aged active region observable in EUV was present on the solar disk. No compact source in HXRs was discernible during the time the FOXSI-3 payload targeted the Sun. We used the entire 6.49 minutes of rocket observation time to assess the HXR quiet Sun flux for this period of minimum solar activity. We set upper limits for this flux implementing two independent techniques. We calculated an upper limit of 9.6$\times10^{-4}$ s$^{-1}$cm$^{-2}$keV$^{-1}$ applying the Li-Ma ON/OFF method over data of one detector flown in FOXSI-2 (for background measurements) and FOXSI-3 (for direct quiet Sun observations). The lowest quiet Sun HXR flux upper limit we report here is 6.0$\times10^{-4}$ s$^{-1}$cm$^{-2}$keV$^{-1}$. We obtained this limit using data from three silicon detectors combined (all flown in FOXSI-3) and applying \citeauthor{gehrels1986confidence}' method, purely based on Poisson statistics.


The exact nature of why the FOXSI-2 upper limit is almost two orders of magnitude higher than the FOXSI-3 limits is not fully clear. Na\"ively, this difference suggests that the quiet Sun HXR flux during a time of intense solar activity (the case for FOXSI-2) might be higher than its counterpart during a minimum in the solar cycle (scenario for FOXSI-3). We can not entirely rule out such a possibility. However, there is a caveat we want to manifest in this case. For FOXSI-2, we isolated a region within the solar disk free of the most intense ghost rays. Yet, some remnant ghost rays from other weaker sources could potentially still be getting into the detectors, affecting our estimated FOXSI-2 constraints. Very recently, \citet{purkhart2022nanoflare} studied nanoflares in quiet Sun regions during solar cycle 24 using SDO/AIA image series to assess their contribution to coronal heating during different levels of solar activity. They reported no correlation between the derived nanoflare energy flux and the solar activity level. Although \citeauthor{purkhart2022nanoflare} analyzes are in EUV and not in HXRs, their results further discourage the hypothesis that the different limits we obtained from FOXSI-2 and -3 reflect an intrinsic disparity in the quiescent emission between different phases of the solar cycle.

Direct focusing HXRs brings the possibility of assessing quiet Sun emission during periods of high solar activity. But, additional optical elements need to be part of the instrument to diminish ghost rays. Future space-based solar HXR telescopes using Wolter-I optics should implement ways to minimize (if not entirely block) ghost rays to analyze quiet Sun emissions during maximums of solar activity. Further observations will give a definite answer on whether or not quiet Sun HXR fluxes correlate with the solar cycle. 

The HXR upper limits we calculate here using FOXSI data can be compared with prior reported constraints. Figure \ref{Fig:hannah} compares our FOXSI limits (in the 5-10 keV energy range) with those estimated by \citet{hannah2010constraining} using almost 12 cumulated days of RHESSI solar off-pointing observations during periods of minimum activity. \citet{hannah2010constraining} binned the limits using the following energy bins; 3-6 keV, 6-12 keV, 12-25 keV, 25-50 keV, 50-100 keV, and 100-200 keV. The upper limits \citet{hannah2010constraining} report for the 3-6 keV and the 6-12 keV energy range, with a $2\sigma$ confidence level, are $3.4\times10^{-2}$ s$^{-1}$cm$^{-2}$keV$^{-1}$ and $9.5\times10^{-4}$ s$^{-1}$cm$^{-2}$keV$^{-1}$, respectively. All the quiet Sun HXR limits we report in this paper are in agreement with those thresholds calculated by \citet{hannah2010constraining}. In particular, the FOXSI-3 limits that correspond to a period of minimum solar activity lie in the same order of magnitude as the 6-12 keV limit from \citet{hannah2010constraining}, $\sim10^{-3}$ s$^{-1}$cm$^{-2}$keV$^{-1}$.

\citet{hannah2010constraining} not only reported upper limits, they also presented interpretations of what these limits imply over possible solar physical processes with the potential of producing HXR emissions. Such interpretations include the assessment of nanoflare isothermal emission, nanoflare non-thermal thick-target and thin-target emissions, and solar Axions. Since the FOXSI limits are not substantially lower than those from \cite{hannah2010constraining}, all their physical interpretations still hold. In particular, FOXSI limits continue to agree with the isothermal emission constraints that \cite{hannah2010constraining} (see Figure 3 in their paper) estimated and compared with results from previous missions that observed the quiet Sun in X-rays, like Sphinx \citep{Sylwester2010}. 

More recently, in \citeyear{marsh2017first}, \citeauthor{marsh2017first} searched for HXR emission in the quiet solar corona with the Nuclear Spectroscopic Telescope Array (NuSTAR) satellite. They used the first observations of the quiet Sun with NuSTAR, which occurred on 2014 November 1. At the time of these observations, an off-axis solar active region contributed a notable amount of ghost rays. \cite{marsh2017first} were interested in searching for transient HXR brightenings present in the quiet Sun. To do so, they looked for increases in HXRs on timescales of 100 s in two energy bands, 2.5-4 keV and 10-20 keV. For the 10-20 keV, they additionally searched brightenings with timescales of 30 and 60 s. \cite{marsh2017first} set upper limits of $\sim$17 s$^{-1}$cm$^{-2}$keV$^{-1}$ for the 2.5-4 keV energy range, and 0.17 s$^{-1}$cm$^{-2}$keV$^{-1}$ for 4-20 keV. This last limit is almost four times higher than the FOXSI-2 upper quiet Sun threshold and two orders of magnitude higher than the limits from FOXSI-3. \cite{marsh2017first} discussed that during their quiet Sun NuSTAR observations, the nonsolar background would be the dominant source of high energy emission in the NuSTAR FoV. \cite{marsh2017first} cited \cite{wik2014nustar}, who give incident background rates of $\sim$2$\times10^{-5}$ s$^{-1}$cm$^{-2}$keV$^{-1}$ for 4-20 keV, to support their argument. 

We highlight that with observations of only $\sim$ one minute for FOXSI-2 and $\sim$ six minutes for FOXSI-3, we obtained quiet Sun HXR upper limits comparable with previous observations $\sim$ 12 days long \citep{hannah2010constraining}. As demonstrated by \cite{Sylwester2010}, with observations extending over months of low levels of solar X-ray activity, the sensitivity increases for quiet Sun emission assessments. Particularly for a satellite mission version of FOXSI, we anticipate a two to three orders of magnitude increased sensitivity. 

\begin{acknowledgements}
The first author of this paper is funded by the NASA FINESST grant 80NSSC19K1438. The FOXSI sounding rocket experiment is funded by NASA grants NNX11AB75G, NNX16AL60G, and 80NSSC21K0030. The Univeristy of Minnesota team is supported by an NSF Faculty Development Grant (AGS-1429512), an NSF CAREER award (NSF-AGS-1752268), and the SolFER DRIVE center (80NSSC20K0627).

This work was also supported by JSPS KAKENHI Grant Numbers JP17H04832, JP16H02170, JP16H03966, JP24244021, JP20244017, and World Premier International Research Center Initiative (WPI), MEXT, Japan.

This work was supported by JSPS KAKENHI Grant Numbers JP18H03724, JP18H05463, JP17H04832, JP16H02170, JP15H03647, JP21540251, JP16H03966, JP24244021, JP20244017, and World Premier International Research Center Initiative (WPI), MEXT, Japan.

The FOXSI team is grateful to the NSROC teams at WSMR and Wallops for the excellent operation of their systems.  Furthermore, the authors would like to acknowledge the contributions of each member of the FOXSI experiment team to the project, particularly our team members at ISAS and Kavli IPMU for the provision of Si and CdTe detectors and at MSFC for the fabrication of the focusing optics.

\end{acknowledgements}

\bibliographystyle{aa}
\bibliography{references}

\end{document}